\documentclass[review]{elsarticle}

\usepackage{lineno}
\usepackage{amsmath}
\usepackage[psamsfonts]{amssymb}
\usepackage[unicode, pdftex]{hyperref}

\modulolinenumbers[5]

\journal{Journal of \LaTeX\ Templates}

\begin{document}

\begin{frontmatter}

\title{Two-Photon Excited Fluorescence Dynamics in Enzyme-Bound NADH:
  the Heterogeneity of Fluorescence Decay Times and Anisotropic Relaxation.}


\author[mymainaddress]{Ioanna A. Gorbunova}

\author[mymainaddress]{Maxim E. Sasin}

\author[mymainaddress]{Alexander A. Semenov}

\author[mymainaddress]{Andrey G. Smolin}

\author[mymainaddress]{Yaroslav M. Beltukov}

\author[mymainaddress,mysecondaryaddress]{Dmitrii P. Golyshev}

\author[mymainaddress]{Oleg S. Vasyutinskii\corref{mycorrespondingauthor}}
\cortext[mycorrespondingauthor]{Corresponding author}
\ead{support@elsevier.com}

\address[mymainaddress]{Ioffe Institute, 26 Polytekhnicheskaya, St.Petersburg, 194021 Russia}
\address[mysecondaryaddress]{Peter the Great St.Petersburg Polytechnic University, 29 Polytechnicheskaya, St.Petersburg, 195251, Russia}

\begin{abstract}
The dynamics of polarized fluorescence in reduced nicotinamide adenine dinucleotide (NADH) at 436~nm under two-photon excitation at 720~nm by femtosecond laser pulses in alcohol dehydrogenase (ADH)-containing buffer solution has been studied experimentally and theoretically. A global fit procedure was used for determination of the fluorescence parameters from experimental data. The interpretation of the experimental results obtained was supported by \emph{ab initio} calculations of NADH structure in solutions. A theoretical model was developed for description of the polarized fluorescence decay in enzyme-NADH binary complexes that considered several possible interaction scenarios. The main results obtained are as follows. We suggest that the origin of a significant enhancement of the nanosecond decay time value in the ADH-bounded NADH compare with the free NADH can be attributed to the significant decrease of non-radiative relaxation probabilities due to decrease of charges separations in the nicotinamide ring in the conditions of an apolar  ADH-NADH binding site environment. The existence of a single decay time in the ADH-NADH complex in the nanosecond time-domain in comparison with two decay times observed in free NADH can be attributed to a single NADH unfolded \emph{trans}-like conformation bounded within the ADH site. Comparison of  the experimental data obtained and the theory developed suggested the existence of an anisotropic relaxation time of about 1 ns related most likely with interactions between excited NADH and the ADH binding site that resulted in the rearrangement of nuclear distribution and rotation of fluorescence transition dipole moment. The analysis of the polarization-insensitive component of the fluorescence decay in ADH-containing solution using multiexponential fitting models suggested that a four-exponential fit provides the best characterization of free and enzyme-bound forms of NADH needed for detailed understanding of the relaxation processes occurring. However, for determination only of the relative concentrations of enzyme-bound and free forms of NADH in solutions and cells that are of the main  importance for biochemical applications, a three-exponential fit provides practically the same level of accuracy.
\end{abstract}

\begin{keyword}
\texttt{elsarticle.cls}\sep \LaTeX\sep Elsevier \sep template
\MSC[2010] 00-01\sep  99-00
\end{keyword}

\end{frontmatter}

\section{Introduction}
\label{sec:Introduction}

Reduced nicotinamide adenine dinucleotide (phosphate) (NAD(P)H)  are intrinsic fluorophores that play crucial role in redox reaction in living cells \cite{Huang2002a,Heikal2010}.  As known \cite{Heikal2010}, NADH and its oxidized form  NAD$^+$ are present in a wide variety of concentrations in living cells and tissues and involved in mitochondrial function, energy metabolism, calcium homeostasis, gene expression, oxidative stress, aging and apoptosis. In their seminal work Chance et al. \cite{Chance1962} have demonstrated that NADH exhibits autofluorescence in its reduced form, whereas NAD$^+$ is not fluorescent.

The  fluorescence studies  of enzyme-NADH complexes have been pioneered by Theorell at al. \cite{Theorell1967} who recognized the binding process of NADH with alcohol dehydrogenase (ADH) as a two-step reaction and concluded that two coenzyme binding sites in ADH are functionally independent. Measured fluorescence intensity in NADH was found to enhance upon coupling with ADH and a blue shift of the ADH-NADH absorption maximum of about 15~nm in comparison with free NADH was observed \cite{Theorell1967,Konig1997}. Theorell et al. \cite{Theorell1971} and Lindman et al. \cite{Lindman1972} also revealed that excitation energy could be transferred from the NADH molecule bound with one ADH subunit to NADH bound with another subunit of the same ADH. However, Gafni et. al. \cite{Gafni1976} suggested that two binding sites in ADH are identical and non-interactive.

Two-photon excited fluorescence of endogenous fluorophores pioneered by Lakowicz latin et al.~\cite{Lakowicz1992} was intensively studied in the last few years \cite{Stringari2017,Huang2002a} as it allows for the measurement of fluorescence lifetimes independently of light intensity and can be used for separation of  fluorophores in their free and protein-bound forms quantitatively.
Fluorescence lifetime imaging microscopy (FLIM) is now widely used for monitoring metabolic pathways activated in normal and pathological cells\cite{Yaseen2017,Evers2018,Schaefer2019}.

Numerous time-resolved fluorescence studies carried out in solutions, cells, and living tissues revealed that after excitation in the first absorption band at 320--370~nm related to the nicotinamide (NA) chromophore group NADH exhibits a multicomponent decay dynamics. In aqueous solution free NADH is known to exhibit biexponential fluorescence decay with the lifetimes of about 0.3~ns and 0.7~ns~\cite{Visser1981,Couprie1994a,Hull2001,Blacker2013,Blacker2019,Gorbunova20b,Gorbunova20c}. In other solvents the lifetimes and corresponding weighting coefficients depend strongly on temperature and solvent type \cite{Ladokhin1995,Couprie1994a,Visser1981,Blacker2013,Gorbunova20c}. At the same time in enzyme-containing solutions and in cells it is most common to fit NADH fluorescence decay in the nanosecond time domain by two exponentials, one of them having the decay times of several ns and another of about 0.4~ns. These two decay times  are usually attributed to the enzyme-bound and free forms, respectively \cite{Gafni1976,Kierdaszuk1996,Konig1997,Niesner2004,Vishwasrao2005,Skala2007,Vergen2012,Sharick2018,Schaefer2019}. However, other authors used multiexponential analysis of the NADH fluorescence decay in enzyme-containing solutions and cells.~\cite{Lakowicz1996,Yu2009,Yaseen2013,Evers2018,Yaseen2017}

In most of the experiments carried out polarization-free (magic angle) geometries were used and fluorescence decay signals  were presented as a sum of several exponentials with corresponding weighting coefficients. However, within this approach it is often inconveniently to assign
multiple exponential fluorescence decays to specific molecular origins \cite{Lakowicz2006}. These problems can be overcome at least partly by using the polarization-sensitive experimental schemes. \cite{Weiner1968,Kierdaszuk1996,Piersma1998}  Vishwasrao et al.  \cite{Vishwasrao2005} and Yu and Heikal \cite{Yu2009} used fluorescence-lifetime imaging microscopy (FLIM) that combined the time-resolved fluorescence and anisotropy measurements and proceeded the global analysis of fluorescence and associated anisotropy decays of intracellular NADH for determination of the relative concentration and free-to-enzyme-bound ratios of this coenzyme that allowed to investigate thin details of the response of NADH to the metabolic transition from normoxia to hypoxia.

Sharick et. al. \cite{Sharick2018} used multiphoton NAD(P)H FLIM for characterisation of the relative concentrations of several metabolic enzymes in solutions and cells and demonstrated that measured NAD(P)H fluorescence  decay times can be used to distinguish NADH bound to different metabolic enzymes in solutions, and thus quantify changes in the relative activities of the enzymes.

Very recently Cao et al.\cite{Cao2020} investigated the ultrafast fluorescence dynamics of NADH-dehydrogenase (MDH/LDH) complexes by using both the femtosecond upconversion and picosecond time-correlated single photon counting (TCSPC) methods and reported the observation of a few-picosecond decay process in bound NADH.

As demonstrated in numerous studies mentioned above, FLIM of NAD(P)H can be used for non-destructive determination of changes in cellular metabolism. However, till now these changes are difficult to interpret \cite{Sharick2018}. The origin of the fluorescence decay times numbers and their values in NADH bound to different enzymes still remains controversial. Therefore, the relative number of NADH molecules bound to different enzymes cannot be obtained without further characterization.  \cite{Cao2020} Moreover, the fluorescence anisotropy and corresponding anisotropic decay times, although very important for quantifying NADH molecules bound to enzymes, are currently mostly used as phenomenological parameters without a well grounded understanding of their mechanisms.

This paper aims to address, at least partly, these important issues. In our recent paper \cite{Gorbunova20c} the decay of polarized fluorescence  in NADH at 460~nm under two-photon excitation at 720~nm was studied in detail in water-methanol solutions as a function of methanol concentration. Based on the analysis of the experimental results obtained and \emph{ab initio} calculations performed we have suggested that the heterogeneity in the measured decay times in free NADH in solutions can be due to different charge distributions in the \emph{cis} and \emph{trans} configurations of the nicotinamide ring that result in different intramolecular electrostatic field distributions and lead to two different non-radiative decay rates. Similar arguments have been suggested \cite{Gorbunova20c} for explanation of the experimental findings of two decay times in mononucleotide  NMNH reported earlier by Krishnamoorthy et al \cite{Krishnamoorthy1987} and Kierdaszuk et al. \cite{Kierdaszuk1996}.

In this paper the approach developed in ref.~\cite{Gorbunova20c} was extended  to enzyme-bounded endogenous fluorophore NADH \emph{in vitro}. The following excited states properties
were studied experimentally and theoretically: a significant enhancement of the observed nanosecond decay time in enzyme-bounded NADH in comparison with its free forms in solution, the heterogeneity of fluorescence decay times in enzyme-bounded NADH as a function of local environment, an accurate determination of free and bound NADH relative concentrations under their simultaneous recording, and the nature of the polarization-sensitive decay times in enzyme-bound NADH.

The decay of polarized fluorescence in NADH-ADH complexes at 436~nm under two-photon excitation at 720~nm in buffer solution was studied experimentally. A global fit procedure was used for determination of the fluorescence parameters from experimental data and a comprehensive analysis of the experimental errors was made. The interpretation of the experimental results obtained was supported by \emph{ab initio} calculations of NADH structure in various solutions. Also, a quasiclassical approach was developed for description of fluorescence decay in enzyme-bound fluorophores that considered several possible interaction scenarios. The conclusions made were compared with the results reported by other authors.

In brief, the main results obtained are as follows.
We suggest that the origin of a significant enhancement of the nanosecond decay time value in the ADH-bounded NADH compare with the free NADH in solution can be attributed to a significant decrease of charges separations in the nicotinamide (NA) ring in the conditions of a typical apolar binding site environment in the ADH-NADH complex.~\cite{Piersma1998,Vishwasrao2005} The charges separations  decrease results in the decrease of non-radiative transition probabilities and in corresponding increase of the decay time observed. Similar reasons are likely responsible for significantly different lifetime values observed in NADH bounded with various enzymes \cite{Sharick2018}. The existence of a single decay time in the ADH-NADH complex in the nanosecond time-domain (contrary to two-exponential decay observed in free NADH in solution~\cite{Visser1981,Couprie1994a,Hull2001,Gorbunova20c}) supports the above model~\cite{Gorbunova20c} because  as known from the X-ray structure spectroscopy and NMR experiments, NAD(P)H imbedded into the ADH binding site is bound in a single unfolded \emph{anti} "\emph{trans}-out-of-plane" nuclear conformation.~\cite{Hammen2002,Plapp17,Vidal2018}

Comparison between the experimental data of the fluorescence anisotropy decay in ADH-bound NADH under two-photon excitation with the results of quasiclassical theory suggests the existence of an isotropic decay time of about 150 ps and an anisotropic decay time of about 1 ns both manifesting different interactions between NADH and ADH. The analysis of the polarization-insensitive component of the fluorescence decay in ADH-containing solution using multiexponential fitting models suggested that a four-exponential fit provides the best characterization of free and enzyme-bound forms of NADH needed for detailed understanding of the relaxation processes occurring. However, for determination only of the relative concentrations of enzyme-bound and free forms of NADH in solutions and cells that are of the main  importance for biochemical applications, a three- and even two-exponential fit can be used without dramatic lost of accuracy.

The organization of the paper is as follows. Section~\ref{sec:methods} contains the description of the experimental method used including experimental data processing and the analysis of experimental errors.  The experimental results obtained and comparison with the results of other authors are given in Sec.~\ref{sec:experimental results}. The results of \emph{ab initio} calculations on NADH optical excitation and structure is given in Sec.~\ref{sec:abinitio}.  Theoretical modelling of the polarized fluorescence decay in ADH-bound NADH is presented in Sec.~\ref{sec:models} The discussions of the obtained results and the models developed are given in Sec.~\ref{sec:Discussion}. The Conclusion summarises the main results obtained.

\section{Materials and methods}
\label{sec:methods}

\subsection{Materials}
Alcohol dehydrogenase equine (lyophilized powder, recombinant, expressed in E. coli, Sigma Aldrich, lot 55689 ) and $\beta$-Nicotinamide adenine dinucleotide (Sigma Aldrigh) were used in the experiment.   All samples were prepared in PBS buffer (10 mM PBS, pH = 7.2) and used without preliminary purification.  For studying of NADH-ADH binary complexes in solution, ADH was dissolved in 1.5 ml PBS in the concentration of 50 $\mu$M. NADH was added to the preliminary prepared ADH solution in the concentration of 25 $\mu$M. An estimated stoichiometric relationship NADH:ADH was 1:2. All solutions were prepared fresh daily at 20$^\circ$C.

\subsection{Time-resolved fluorescence measurements}
\label{sec:time_resolved}
The experimental setup used was similar to that described in detail in our previous publications.~\cite{Denicke10,Herbrich15,Sasin18,Sasin19}
Briefly, two-photon excitation of ADH-contained solutions at 720 nm was performed by a linearly polarized laser beam. A femtosecond Ti:Sapphire oscillator (Mai Tai HP DS, Spectra Physics) tunable in the spectral range of 690 -- 1040~nm with a pulse duration of 100~fs and a repetition rate of 80.4~MHz was used as an excitation source. The laser beam polarization was controlled by a half waveplate and the  polarization degree was better than 0.995. The laser beam was expanded by a telescope to a diameter of 4~mm and then focused onto the center of a quartz cuvette with ADH-containing solution. Average laser beam intensity on the cuvette was about 500~mW.

The fluorescence was collected in the direction perpendicular to the laser beam.  The  polarization components of the fluorescence intensity parallel $I_{\parallel}$ and perpendicular $I_{\perp}$ with respect to the excitation polarization plane were separated by a Glan  prism and then simultaneously detected by two ultrafast avalanche photodetectors (SPAD, ADP-050-CTC, MPD) operated in a photon-counting mode.  To achieve the maximum possible selectivity of NADH isolation  the fluorescence intensity was detected within the spectral range of 426 -- 446~nm  selected by an interference filter  ZET436/20x (Chroma). The fluorescence spectral range was  located in close proximity of the ADH-bound NADH absorption maximum~\cite{Huang2002a}.  The signals were analyzed by a TCSPC system (PicoHarp 300, PicoQuant) and were typically collected during 30 min with a time bin of 4~ps.

\subsection{Experimental data processing}
The recorded fluorescence orthogonal polarization components $I_{\parallel}$(t) and $I_{\perp}$(t) can be presented by the expressions:~\cite{Denicke10,Sasin19}
\begin{equation}
\label{funcy}
 I_{\parallel}(t)= G\int\limits_0^{t} \mathrm{IRF}({t'})I_{l}(t-{t'})[1+2r_l (t-{t'})]d{t'}
\end{equation}
\begin{equation}
\label{funcx}
 I_{\perp}(t)= \int\limits_0^{t}  \mathrm{IRF}({t'})I_{l}(t-{t'})[1-r_l (t-{t'})]d{t'},
\end{equation}
where $I_{l}(t-{t'})$ and $r_l(t-{t'})$ are isotropic intensity and anisotropy, respectively, the subscript index $l$ refers to the linear polarization of the excitation laser beam, IRF(t) is an instrumental response function, and G is a ratio of sensitivity of fluorescence detection channels.

The polarization-insensitive fluorescence intensity $I_{tot}(t)$  and the anisotropy $r(t)$ can be calculated from the experimentally determined polarization components $I_{\parallel}$(t) and $I_{\perp}$(t) in eqs.~(\ref{funcy}) and (\ref{funcx}) according to~\cite{Lakowicz2006}:
\begin{equation}
\label{iso_func}
 I_{tot}(t)= \frac{\frac{1}{G}I_\parallel(t)+2I_\perp(t)}{3},
\end{equation}
\begin{equation}
\label{aniso_func}
 r(t)= \frac{\frac{1}{G}I_\parallel(t)-I_\perp(t)}{\frac{1}{G}I_\parallel(t)+2I_\perp(t)}.
\end{equation}

The function $I_{tot}(t)$ is a measure of the population of the excited state and is usually presented as a sum of exponentials~\cite{Visser1981,Ladokhin1995,Couprie1994a,Hull2001,Vishwasrao2005,Blacker2013,Blacker2019,Gorbunova20c}:
\begin{equation}
\label{sum_exp}
 I_{tot}(t)= I_0\sum\limits_{i=1}^na_{i}\,\exp\left(-\frac{t}{\tau_{i}}\right),
\end{equation}
where $I_0$ is a time-independent initial fluorescence intensity, $\tau_i$ are decay times, and $a_i$ are corresponding weighting coefficients that are normalized to unity: $\sum_i a_i=1$.

The anisotropy $r(t)$ cannot in general be presented as a sum of several exponentials because the time-dependent terms in the denominator of eq.~(\ref{aniso_func}) cannot be cancelled out (see e.g. discussion in ref.~\cite{Vishwasrao2005}). Therefore, we did not fit the anisotropy $r(t)$ directly. Instead, the global analysis~\cite{Lakowicz2006} of eqs.~(\ref{funcy}) and (\ref{funcx}) was proceeded where the fluorescence decay times $\tau_i$, weighting coefficients $a_i$, rotational diffusion times $\tau_{ri}$, and anisotropies $r_i$ were used as fitting parameters.

The fitting procedure and errors estimation were similar to that described in detail in our recent publication~\cite{Gorbunova20c}. As in the conditions of our experiments the number of photon counts was relatively small, the maximum likelihood method with Poisson likelihood function~\cite{Hauschild2001,Laurence2010} $\chi^2$ was used in the global fit procedure. Minimization of the cost function was carried out by stochastic optimization algorithm using differential evolution implemented in \textit{SciPy}.


\subsection{Instrumental response function of SPAD photodetectors}
\begin{figure} [ht]
   \begin{center}
   \begin{tabular}{c}
   \includegraphics[height=8cm]{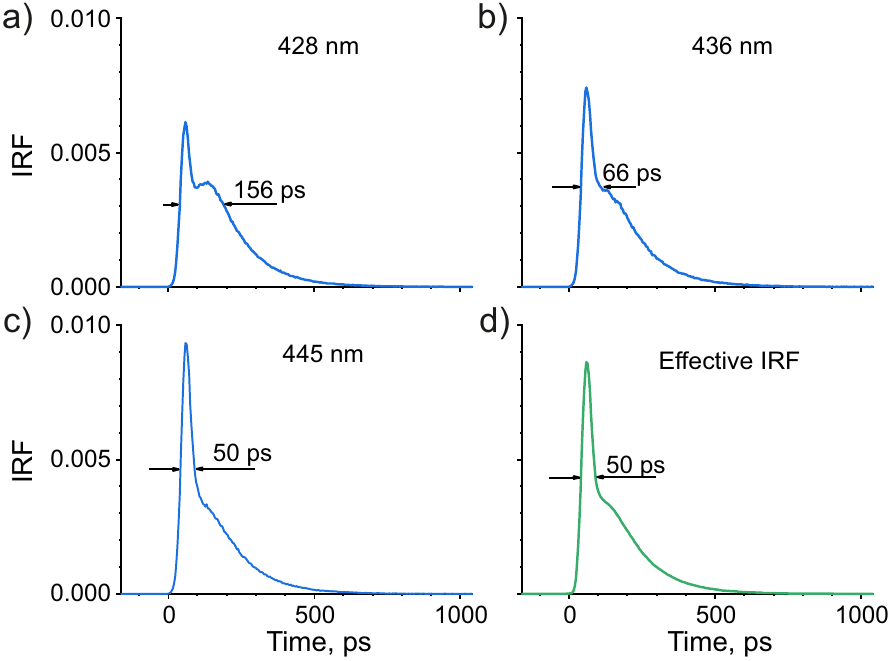}
   \end{tabular}
   \end{center}
   \caption{Instrumental response functions of SPAD photodetectors. \\
   Experimental data determined at wavelengths: a) 428 nm, b) 436 nm, c) 445 nm. \\
    d) Calculated effective IRF in the wavelength region of 426--446~nm.}
    \label{fig:irf}
\end{figure}

In the case of molecular sample excitation with femtosecond laser pulses the temporal resolution of recorded fluorescence signals is usually governed by a photodetector's timing resolution. As known~\cite{Giudice2007} the IRF of SPAD photodetectors used in our experiments is strongly wavelength-dependent. As the IRF shape was critical for accurate analysis of the experimental data, it was carefully determined and analysed as a function of the detected light wavelength before experiments commencement. The IRFs  were determined experimentally by recording the femtosecond laser pulses scattered from a matt metallic surface.  Typical IRFs obtained at three wavelengths in the spectral range of 428--445~nm are shown in Fig.~\ref{fig:irf}a--c and the effective IRF calculated as explained below is shown in Fig.~\ref{fig:irf}d.

As can be seen in Fig.~\ref{fig:irf}a--c in the wavelength region of 428--445~nm IRF has a complex double-peak shape that depends strongly on the particular wavelength. For increasing the accuracy of the fluorescence decay times determination in the 436$\pm$10~nm spectral region an effective IRF was built as a sum of seven IRFs determined each at a definite wavelength in this spectral range with appropriate weighting coefficients.  The weighting coefficients were calculated by fit of the fluorescence decay signals in NADH in aqueous solution using eqs.~(\ref{funcy})--(\ref{iso_func}), and (\ref{sum_exp}) with fixed fluorescence parameters $G$, $\tau_i$, $a_i$ determined in our recent paper~\cite{Gorbunova20c}.
The obtained effective IRF used for proceeding of all experimental data recorded in ADH-containing solutions is shown in Fig.~\ref{fig:irf}d and in inset in Fig.~\ref{yx_curves}.


\section{Experimental results}
\label{sec:experimental results}
\subsection{Polarized fluorescence decay in binary complex ADH-NADH in solution}
\begin{figure} [ht]
   \begin{center}
   \begin{tabular}{c} 
   \includegraphics[height=9cm]{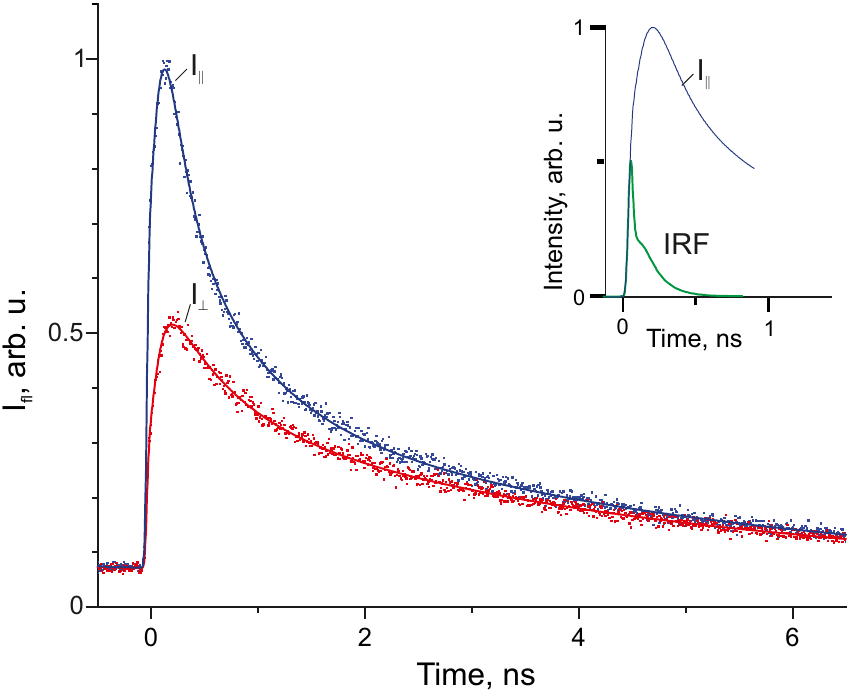}
   \end{tabular}
   \end{center}
   \caption[]
   {  Two orthogonally polarized fluorescence components $I_\parallel(t)$ (blue dots) and $I_\perp(t)$ (red dots) in NADH-ADH binary complexes in solution. The blue and red solid lines are the best fits.  IRF(t) (green solid line) and rising edge of $I_\parallel(t)$ (blue solid line) are shown in inset.}
   \label{yx_curves}
\end{figure}

Polarized fluorescence decay in NADH-ADH binary complexes was recorded in  50 $\mu$M ADH solution with NADH added at the concentration of 25~$\mu$M. Two typical orthogonally polarized fluorescence components $I_\parallel(t)$  and $I_\perp(t)$ are shown in Fig. \ref{yx_curves}.

The polarized fluorescence decay signals in Fig.~\ref{yx_curves} exhibit a rising edge that practically follows the edge of IRF(t) and a slowly decaying tail.  Two orthogonally polarized fluorescence decay components were analyzed by a global fit procedure described in Sec.~\ref{sec:methods} above. The polarization-insensitive fluorescence intensity $I_{tot}(t)$  was determined from the experimental signals in Fig.~\ref{yx_curves} using eq.~(\ref{iso_func}) and then analyzed using the multiexponential function in eq.~(\ref{sum_exp}). The analysis of different multiexponential models and the decay times obtained is given in Sec.~\ref{sec:isotropic}.

 \subsection{Fluorescence anisotropy}
\begin{figure} [h]
   \begin{center}
   \begin{tabular}{c}
   \includegraphics[height=7cm]{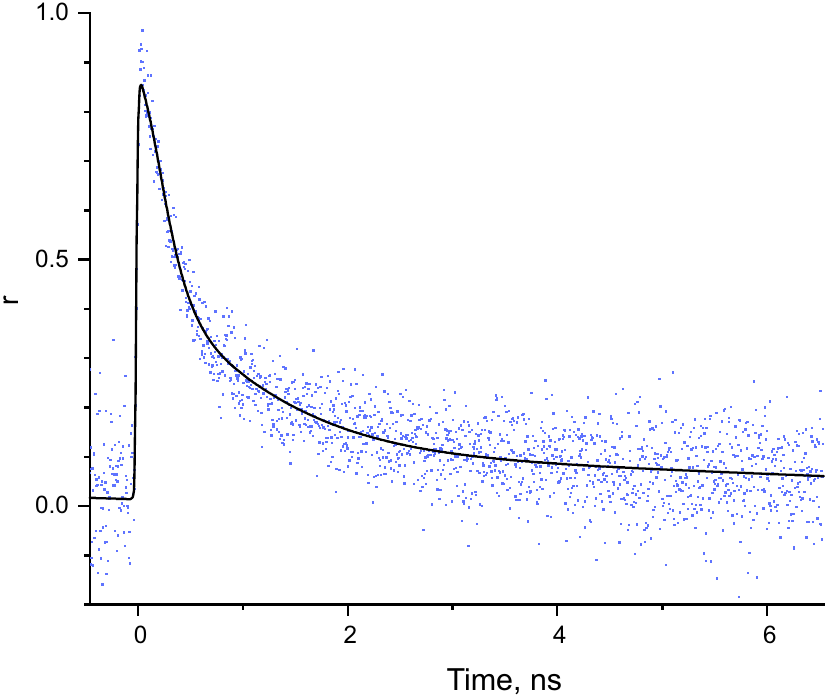}
   \end{tabular}
   \end{center}
   \caption{The fluorescence anisotropy in NADH-ADH binary complex in solution. \\
   Dots are experimental data and the solid line was calculated using the fitting curves obtained for the orthogonal polarization components $I_\parallel(t)$ and $I_\perp(t)$ according to eq.~(\ref{aniso_func}). }
    \label{fig:anisotropy}
\end{figure}

The fluorescence anisotropy $r(t)$ obtained from the experimental signals in Fig.~\ref{yx_curves} using  eq.~(\ref{aniso_func}) is given in Fig.~\ref{fig:anisotropy}.

The fluorescence anisotropy in enzyme-bound NADH was earlier studied by Piersma et al.~\cite{Piersma1998} and by Vishwasrao et al.~\cite{Vishwasrao2005}  Piersma et al.~\cite{Piersma1998} studied the anisotropy decay
of the enzyme-bound NADH in enzyme complexes np-ADH and ADH and found that NADH was fully immobilized inside both of them and that no fast rotations were observed in the timescale of 0.1-5~ns. This result was supported by Vishwasrao et al.~\cite{Vishwasrao2005} who studied the fluorescence anisotropy of NADH bound to m-malate dehydrogenase (mMDH) and reported that NADH was rigidly bound without significant segmental mobility
within the enzyme.

The anisotropy $r(t)$ in Fig.~\ref{fig:anisotropy} contains the contributions from free and ADH-bound forms of NADH. The analysis of these contributions and comparison with earlier studies are given in Sec.~\ref{sec:Discussion}.

\section{\emph{Ab initio} calculations}
\label{sec:abinitio}
\emph{Ab initio} calculations of the structure, vertical excitation energies, and transition dipole moments of twenty four  conformers of NADH in water and methanol solutions in the ground and excited states have been reported in our recent publication~\cite{Gorbunova20c}.

In this paper electronic structure computations have been performed for a NADH conformation with the nuclear configuration related to ADH-bound NADH.~
This nuclear configuration was   documented precisely from the X-ray structure spectroscopy and NMR experiments, it is known to belong to unfolded \emph{trans}-like \emph{anti}  conformation~\cite{Vidal2018,Plapp17}. The \emph{ab initio} calculations were carried out by means of the polarizable continuum model (PCM) at the B3LYP-D3BJ/6-31G* level with the GAUSSIAN package~\cite{GAUSSIAN09}. The functionals were extended with the D3 version of Grimme's dispersion correction.~\cite{Kovacs2017} The calculations were performed in water and other solutions with various polarities at the fixed NADH nuclear configuration with different solutions modelled different cite polarities.

\begin{figure} [h]
   \begin{center}
   \begin{tabular}{c}
   \includegraphics[height=7cm]{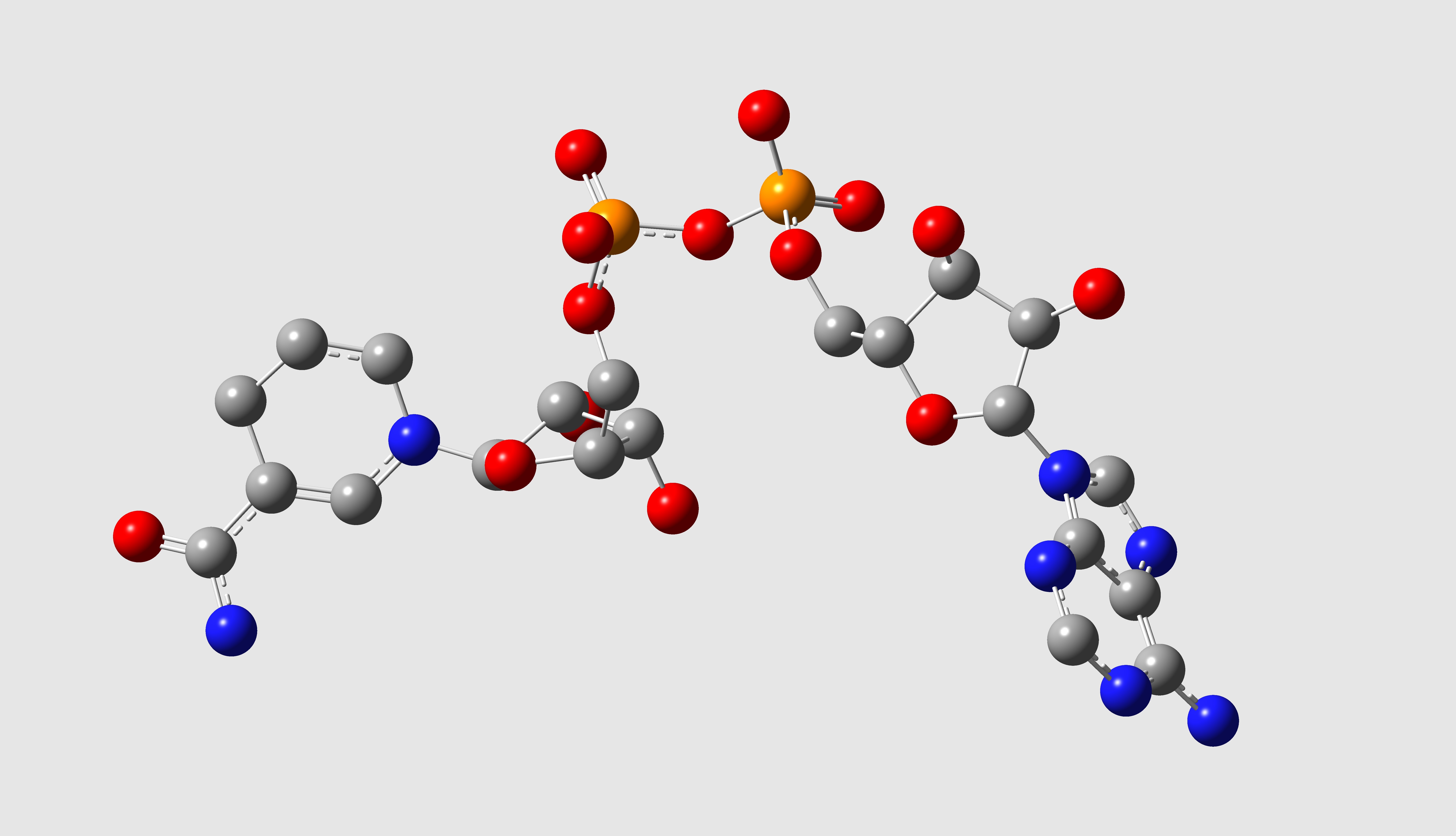}
   \end{tabular}
   \end{center}
   \caption{A schematic of the  ADH-bound NADH conformation possessing the unfolded \emph{trans}-like \emph{anti} nuclear configuration. The grey, red, blue,  and orange balls represent carbon, oxygen, nitrogen, and phosphorus atoms, respectively.}
    \label{fig:ADH-NADH}
\end{figure}

A   schematic of the NADH conformation under study  is shown in Fig.~\ref{fig:ADH-NADH} (PDB ID: 4XD2). As known \cite{Vidal2018} ADH in cells exists as a dimer that can contain two NADH molecules simultaneously. According to the results of X-ray structure spectroscopy \cite{Plapp17} and our  \emph{ab initio} calculations the directions of the transition dipole moments of the two NADH molecules are almost perpendicular to each other.

\section{Theoretical modelling of the polarized fluorescence decay in ADH-bound NADH under two-photon excitation}
\label{sec:models}
We consider polarized fluorescence decay from two identical non-rotating coenzyme molecular probes bound with an enzyme dimer and excited by a short laser pulse. The probes are assumed to be bounded within an enzyme binding site each and fixed in the body frame. The probes can transfer energy between each other and also interact with non-fluorescing environment within the site~\cite{Zhang2015,Vidal2018}. The expressions describing the fluorescence decay in the case of a one-photon excitation were reported  by Tanaka et al.~\cite{Tanaka-1979}. The two-photon excitation case is considered below.

\subsection{Energy transfer}
Considering the incoherent energy transfer (FRET)~\cite{Clegg-2010} between two coenzyme molecules 1 and 2, the evolution of the corresponding excited state populations $n_1$, $n_2$ can be described by the following set of master equations:~\cite{Tanaka-1979}

\begin{eqnarray}
\label{eq:master}
	\frac{dn_1(t)}{dt} = -\gamma_1 n_1(t) - W_{12} n_1(t) + W_{21} n_2(t),
	\\
	\frac{dn_2(t)}{dt} = -\gamma_2 n_2(t) - W_{21} n_2(t) + W_{12} n_1(t),
\label{eq:master1}
\end{eqnarray}

where $\gamma_1$ and $\gamma_2$ are decay rates of molecules 1 and 2, respectively and the energy transfer is described by exchange rates $W_{12}$ and $W_{21}$.

Each decay rate $\gamma_i$ in eqs.~(\ref{eq:master}) and (\ref{eq:master1}) can be a sum of radiation and non-radiation decay rates: $\gamma_i=\gamma_i^{\rm rad}+\gamma_i^{\rm nrad}$, where the non-radiation decay rate $\gamma_i^{\rm nrad}$ can be especially affected by interaction with microenvironment. Assuming that the excitation laser pulse applied to the molecular system at the time $t=0$ is much shorter than the characteristic evolution time of the populations  $n_1(t)$ and $n_2(t)$,  the solution of the set of equations (\ref{eq:master}), (\ref{eq:master1}) can be written in the form:
\begin{eqnarray}
	n_1(t) = n_1(0) f_{11}(t) + n_2(0) f_{21}(t),
	\\
	n_2(t) = n_1(0) f_{12}(t) + n_2(0) f_{22}(t),
\end{eqnarray}
where $n_1(0)$ and $n_2(0)$ are populations at $t=0$ and
\begin{eqnarray}
f_{11}(t) &=& \alpha_1 e^{-\varkappa_1 t} +  \alpha_2 e^{-\varkappa_2 t}, \\
f_{12}(t) &=& \frac{W_{12}}{\Delta} \left(-e^{-\varkappa_1 t} + e^{-\varkappa_2 t}\right), \\
f_{21}(t) &=& \frac{W_{21}}{\Delta} \left(-e^{-\varkappa_1 t} + e^{-\varkappa_2 t}\right), \\
f_{22}(t) &=& \alpha_2 e^{-\varkappa_1 t} +  \alpha_1 e^{-\varkappa_2 t},
\end{eqnarray}
where
\begin{align}
	\varkappa_{1,2} &= \frac{\gamma_1 + \gamma_2 + W_{12} + W_{21}}{2} \pm \frac{\Delta}{2}, \label{eq:kappa}\\
	\alpha_{1,2} &= \frac{1}{2} \pm \frac{\gamma_1 - \gamma_2 + W_{12} - W_{21}}{2\Delta}, \\
	\Delta^2 &= (\gamma_1+\gamma_2-W_{12} - W_{21})^2 \notag\\
	& \qquad  -4 (\gamma_1 \gamma_2 - \gamma_1 W_{12} - \gamma_2 W_{21}). \label{eq:Delta}	
\end{align}

In the case of instant two-photon excitation of the molecules 1 and 2 the excited state populations $n_1(0)$ and $n_2(0)$  are proportional to the square of the corresponding interaction energy~\cite{Lakowicz2006}.
\begin{equation}
	n_i(0) \propto \big|\mathbf{e}_{\rm pu} \cdot \mathbf{S}_{i}^{\rm lab} \cdot \mathbf{e}_{\rm pu}\big|^2,
\end{equation}
where $\mathbf{e}_{\rm pu}$ is an excitation light polarization vector and $\mathbf{S}_{i}^{\rm lab}$ is a two-photon absorption tensor written in the laboratory frame. The tensor $\mathbf{S}_{i}^{\rm lab}$ can be transformed to the body frame according to: $\mathbf{S}_{i}^{\rm lab}={\mathbf{R} \cdot \mathbf{S}_{i} \cdot \mathbf{R}^T}$, where $\mathbf{R}$ is the rotation matrix~\cite{Tanaka-1979} and $\mathbf{S}_{i}$ is the  two-photon absorption tensor written in the body-frame.

\subsection{Molecular fluorescence intensity}
The molecular fluorescence intensity emitted by the molecules 1 and 2 at the time $t$ after the laser pulse can be written in the form:
\begin{equation}
\label{eq:I(t)}
	I(t) = \sum_{j=1,2}\left\langle n_j(t) \big|\mathbf{d}_{j}^{\rm lab}(t) \cdot \mathbf{e}_{\rm fl}\big|^2 \right\rangle_{\mathbf{R}},
\end{equation}
where $\mathbf{e}_{\rm fl}$ is a fluorescence light polarization vector and $\mathbf{d}_{j}^{\rm lab}(t) = \mathbf{R} \cdot \mathbf{d}_{j}(t)$ is the fluorescence transition dipole moment (FTDM).

In the body frame the FTDM $ \mathbf{d}_{j}(t)$  in eq.~(\ref{eq:I(t)}) can in general depend on time $t$ due to intramolecular interactions and interactions with environment resulting in anisotropic vibrational relaxation. These interactions lead to the change of the FTDM value and to the  FTDM rotation in the body frame because of the restructurisation of the excited state nuclear configuration.

Angular brackets in eq.~(\ref{eq:I(t)}) denote averaging over all possible orientations $\mathbf{R}$. Assuming an isotropic distribution of initial orientations of the enzyme, the fluorescence intensity can be presented in the form:
\begin{equation}
\label{eq:Itot}
	I(t) = I_1(t) + I_2(t)
	= J_{11}(t) f_{11}(t) + J_{12}(t) f_{12}(t)
	+ J_{21}(t) f_{21}(t) + J_{22} (t)f_{22}(t),
\end{equation}
where
\begin{equation}
J_{ij}(t) = \Big\langle \big|\mathbf{e}_{\rm pu} \cdot \mathbf{S}_{i}^{\rm lab} \cdot \mathbf{e}_{\rm pu}\big|^2 \big|\mathbf{d}_{j}^{\rm lab}(t) \cdot \mathbf{e}_{\rm fl}\big|^2 \Big\rangle_{\mathbf{R}}.   \label{eq:aver}
\end{equation}

Four terms in eq.~(\ref{eq:Itot}) have clear physical meanings describing the contributions to the fluorescence intensity from the excitation of the molecule $i$ and following emission of the molecule $j$, where $i,j = 1,2$.  Orientational averaging in eq.~(\ref{eq:aver}) for arbitrary pump and fluorescence light polarizations was performed by McClain~\cite{McClain1972,McClain1973}. Similar expressions based on the spherical tensor approach were reported by Vasyutinskii et al.~\cite{Denicke10,Shternin10} Here we consider linear light polarizations only. Then, both polarization vectors $\mathbf{e}_{\rm pu}$ and $\mathbf{e}_{\rm fl}$ are real. If the optical transitions occur between non-degenerate molecular electronic states the two-photon excitation tensors $\mathbf{S}_i$ and the fluorescence transition dipole moments $\mathbf{d}_j(t)$ in eq.~(\ref{eq:aver}) are also real~\cite{McClain1972}. In this case the orientational averaging in eq.~(\ref{eq:aver}) can be presented in the following simple form:
\begin{equation}
\label{eq:Jij}
	J_{ij}(t) = \frac{T_i d_{j}^{\,2}(t)}{45}\Big[1 + 2P_2(\cos\varphi)r_{ij}(t)\Big],
\end{equation}
where $\varphi$ is an angle between the light polarization vectors $\mathbf{e}_{\rm pu}$ and $\mathbf{e}_{\rm fl}$, $P_2(x)=(3x^2 - 1)/2$ is the second order Legendre polynomial, and $d_{j}^{\,2}(t)=\mathbf{d}_{j}(t)\cdot\mathbf{d}_{j}(t)$.

The two-photon anisotropy $r_{ij}(t)$ in eq.~(\ref{eq:Jij}) is given by:
\begin{equation}
	r_{ij}(t) = \frac{2}{7} \left(3\frac{D_{ij}(t)}{T_id_j^{\,2}(t)} - 1 \right),   \label{eq:r}
\end{equation}
where
\begin{equation}
\label{eq:Dij}
D_{ij}(t) = 2 \Big(\mathbf{d}_{j}(t)  \cdot \mathbf{S}_{i}^2  \cdot  \mathbf{d}_{j}(t)\Big) + \Big(\mathbf{d}_{j}(t)  \cdot  \mathbf{S}_{i}  \cdot  \mathbf{d}_{j}(t)\Big)\text{Tr} \mathbf{S}_{i},
\end{equation}
and
\begin{equation}
\label{eq:T_i}
T_i = \text{Tr}^2 \mathbf{S}_{i} + 2 \text{Tr}\mathbf{S}_{i}^2.
\end{equation}

If one of the principal values of the two-photon absorption tensor $\mathbf{S}_i$  is much larger than the others, the tensor can be presented in a dyadic form $\mathbf{S}_{i}=S(\mathbf{p}_{i}\otimes \mathbf{p}_{i})$, where $\mathbf{p}_i$ is the unit vector directed along the axis related to this principal value and $S$ is a constant. Then, the anisotropy $r_{ij}(t)$ in eq.~(\ref{eq:r}) can be transformed to a simple form:~\cite{Chen1993}
\begin{equation}
	r_{ij}(t) = \frac{4}{7}P_2(\cos\theta_{ij}(t)),   \label{eq:r_dyad}
\end{equation}
where $\theta_{ij}(t)$ is the angle between the vectors $\mathbf{p}_{i}$ and $\mathbf{d}_{j}(t)$.

The two-photon anisotropy $r_{ij}(t)$ in eq.~(\ref{eq:r_dyad}) has almost the same form as the one-photon anisotropy~\cite{Tanaka-1979}, but with the extremum values of $4/7$ and $-2/7$~\cite{Callis2002}. Note that in general the two-photon anisotropy $r_{ij}(t)$ should be written in the form of eq.~(\ref{eq:r}) having the extremum values of $(1 \pm 3\sqrt{6/5})/7$ \cite{Callis1993}.

Assuming that the vibrational relaxation can be characterised by a single vibrational relaxation time $\tau_v$ each partial anisotropy $r_{ij}(t)$ in eq.~(\ref{eq:r}) can be presented as~\cite{Gorbunova20b}:
\begin{equation}
\label{eq:rij}
	r_{ij}(t) = \Big[r^{(e)}_{ij} + \big(r^{(g)}_{ij} - r^{(e)}_{ij}\big)e^{-t/\tau_v}\Big]e^{-t/\tau_r},
\end{equation}
where $\tau_r$ is a rotational diffusion time~\cite{Lakowicz2006} and $\tau_v$ is an anisotropic vibrational relaxation time~\cite{Gorbunova20b}.

The terms $r^{(g)}_{ij}$ and $r^{(e)}_{ij}$ in eq.~(\ref{eq:rij}) have clear physical meanings being the initial ground state and relaxed excited state partial molecular anisotropies. They depend on the corresponding NADH nuclear configurations and in general can be either determined from experiment, or calculated from theory. The rotational diffusion time $\tau_r$ in eq.~(\ref{eq:rij}) describes the rotation of the protein as a whole with embedded fluorescence molecular probes.

If the excitation laser beam is polarized along laboratory axis Z, the fluorescence intensity polarization components $I_\parallel(t)$ and $I_\perp(t)$ should be calculated from eq.~(\ref{eq:Jij}) with the angles $\varphi=0$ and $\varphi=\pi/2$, respectively. For arbitrary values of the reaction rates $\gamma_1$,  $\gamma_2$,  $W_{12}$, and $W_{21}$ in eqs.~(\ref{eq:master}), (\ref{eq:master1})  the total anisotropy $r(t)$ in eq.~(\ref{aniso_func}) depends on the values $r_{ij}(t)$ in eq.(\ref{eq:r}) where $i,j=1,2$ in a complicated way. Three important particular cases of the fluorescence intensity $I(t)$  are discussed below.

\subsection{Model I: Energy exchange between the enzyme-bound fluorescence probe and a non-fluorescing amino acid }
\label{sec:nonfluores}
Here we consider the fluorescence decay in presence of a reversible reaction between an enzyme-bound fluorescence probe and a non-fluorescing amino acides in the enzyme binding site. If the absorption bands of the fluorescence probe and the amino acid do not overlap the terms $\mathbf{d}_2$ and $\mathbf{S}_2$ in eqs.~(\ref{eq:aver})--(\ref{eq:T_i}) can be set to zero. In this case the intensity $I(t)$ in eq.~(\ref{eq:Itot}) can be written in the form:
\begin{equation}
\label{eq:I(t)nsym}
	I(t) = I_0 \Big(a_1 e^{-\varkappa_1 t} + a_2 e^{-\varkappa_2 t}\Big)\big[1 + 2P_2(\cos\varphi)r(t)\big],
\end{equation}
where
\begin{align}
	I_0 &= \frac{T_1d_1^{\,2}}{45},\\
	r(t) &= r_{11}(t).
\label{eq:r11}
\end{align}

As can be seen the total fluorescence intensity $I(t)$ in eq.~(\ref{eq:I(t)nsym}) contains two isotropic decay rates $\varkappa_1$ and $\varkappa_2$ that are defined in eq.~(\ref{eq:kappa}) and depends on radiative and nonradiative fluorescence decays in the molecular probe and energy exchange rate  between the molecular probe and amino acid. Two different decay rates $\varkappa_1$ and $\varkappa_2$ appear due to the reversible reaction between the enzyme-bounded fluorescence probe and a non-fluorescing amino acid. The appearance of an additional decay time due to reversible reaction in the excited state   is known and well documented experimentally and theoretically.~\cite{Gafni1976,Ladokhin1995} Note, that in general there can be another additional time-dependent contribution to the fluorescence signal coming from the change of the FTDM $d$ value in eq.~(\ref{eq:I0}) that was already discussed in sec.~\ref{sec:identical} above and assumed to be small.

According to eqs.~(\ref{eq:rij}) and (\ref{eq:I(t)nsym}) in the presence of a reversible reaction with non-fluorescing molecule the fluorescence intensity $I(t)$ depends on two isotropic decay rates: $\varkappa_1$ and $\varkappa_2$, and two anisotropic decay rates: $\gamma_v=\tau_v^{-1}$ and $\gamma_r=\tau^{-1}_{r}$, the former describes fast vibrational relaxation due to the restructuring of the excited state nuclear configuration resulting in rotation of the FTDM and the latter describes the rotational diffusion of the whole enzyme.

\subsection{Model II: Energy exchange between two identical enzyme-bound fluorescence probes}
\label{sec:identical}
In this case the reaction rates related to each molecular probe in eqs.~(\ref{eq:master}), (\ref{eq:master1}), (\ref{eq:kappa})--(\ref{eq:Delta}) can be taken as: $W_{12} = W_{21} = W$, $\gamma_1 = \gamma_2 = \gamma$, $T_i=T$, $\varkappa_1 = \gamma + 2W$, $\varkappa_2 = \gamma$. Then, the fluorescence intensity $I(t)$ in eq.~(\ref{eq:Itot}) can be presented in the  form:
\begin{equation}
\label{eq:I(t)sym}
	I(t) = I_0 e^{-\gamma t} \big[1 + 2P_2(\cos\varphi)r(t)\big],
\end{equation}
where
\begin{equation}
\label{eq:I0}
	I_0 = \frac{2 T d^{\,2}}{45}
\end{equation}
and
\begin{equation}
\label{eq:r(t)sym}
	r(t) = \left[\frac{r_{11}(t) + r_{12}(t)}{2} + \frac{r_{11}(t) - r_{12}(t)}{2}\,e^{-2Wt}\right]e^{-t/\tau_r},
\end{equation}
where the partial anisotropies $r_{ij}$ with $i,j=1,2$ are given in eq.~(\ref{eq:rij}).

Equation~(\ref{eq:I(t)sym}) describes the fluorescence decay in the presence of a reversible reaction between two identical fluorescence probes. According to eqs.~(\ref{eq:rij}) and (\ref{eq:I(t)sym})--(\ref{eq:r(t)sym}) the fluorescence intensity $I(t)$ contains a single polarization-insensitive  (isotropic) decay rate $\gamma$ and three polarization-sensitive (anisotropic) decay rates: $\gamma_v=\tau_v^{-1}$, $2W$, and $\gamma_r=\tau^{-1}_r$.

The isotropic decay rate $\gamma$  describes fluorescence radiative decay and nonradiative  decay due to interaction of each molecular probe with corresponding site environment. Note that the FTDM $d$ in eq.~(\ref{eq:I0}) can also depend on time due to fast restructuring of the molecular probe nuclear configuration in the excited state. However the contribution from this mechanism to the fluorescence intensity is likely not large. Anyway, both effects are isotropic and can be canceled through measuring of the fluorescence anisotropy $r(t)$ in eq.~(\ref{aniso_func}).

According to eqs.~(\ref{eq:rij}) and (\ref{eq:r(t)sym}) even in the simplest case the fluorescence intensity $I(t)$ can contain two anisotropic decay rates, $\gamma_v=\tau_v^{-1}$ and $2W$. The first decay rate $\gamma_v$ describes vibrational relaxation in the course of restructurization of the excited state nuclear configuration and results in rotation of the fluorescence probe FTDM. Another decay rate, $2W$, is the twice of the exchange rate between the fluorescence probes.

\subsection{Model III: No energy exchange between the enzyme-bound fluorescence probe and the enzyme binding site }
\label{sec:noexchange}

In this case the fluorescence intensity $I(t)$ in eq.~(\ref{eq:Itot}) can be presented in the form of eq.~(\ref{eq:I(t)sym}) with the anisotropy $r(t)$ given in eq.~(\ref{eq:r11}). Therefore, if no exchange reaction occurs the fluorescence intensity $I(t)$ depends on a single isotropic decay rate $\gamma$ and a single anisotropic decay rate, $\gamma_v=\tau_v^{-1}$. This case is a particular case of that given in Sec.~\ref{sec:identical}, with $W=0$.

\begin{figure} [h]
   \begin{center}
   \begin{tabular}{c}
   \includegraphics[height=7cm]{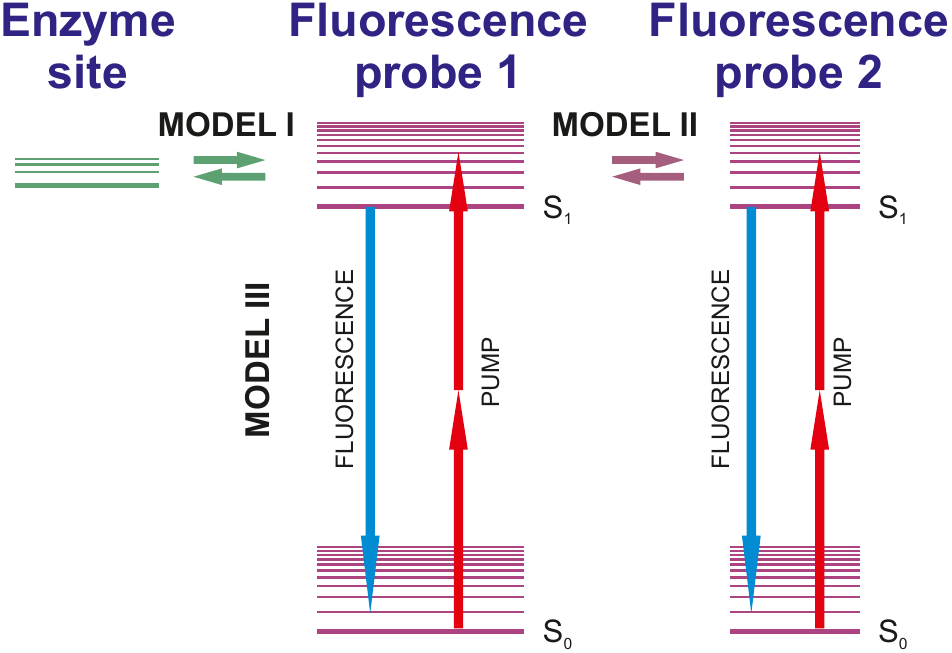}
   \end{tabular}
   \end{center}
   \caption{Schematic of three models used.\\
   Model I: Energy exchange between the enzyme-bound fluorescence probe and a non-fluorescing amino acid (Sec.~\ref{sec:nonfluores}). \\
   Model II: Energy exchange between two identical enzyme-bound fluorescence probes (Sec.~\ref{sec:identical}).\\
      Model III: No energy exchange between the enzyme-bound fluorescence probe and the enzyme bounding site (Sec.~\ref{sec:noexchange}).  }
    \label{fig:Scheme}
\end{figure}

All three models considered are schematically shown in Fig.~\ref{fig:Scheme}.

\newpage

\section{Discussion}
\label{sec:Discussion}
\subsection{A polarization-insensitive component of fluorescence signals and relative concentrations of free and ADH-bound NADH}
\label{sec:isotropic}

As already mentioned in the introduction, the fluorescence lifetime imaging technique is known as a powerful tool for distinction between free and bound NADH fluorescence in cells~\cite{Lakowicz1996,Vishwasrao2005,Sharick2018,Niesner2004} because the lifetime in enzyme-bound NADH enhances
significantly (up to 10 times) compare to that in free NADH. However, the fluorescence decay of bound NADH is usually multi exponential
with shorter components that can be comparable with the decay time of free NADH~\cite{Vishwasrao2005}. Therefore, the interpretation of the underlaying energy transfer mechanisms and determination of the relative concentrations of free and enzyme-bound NADH that despite of many studies done still remained controversial.

\begin{figure} [ht]
   \begin{center}
   \begin{tabular}{c} 
   \includegraphics[height=10cm]{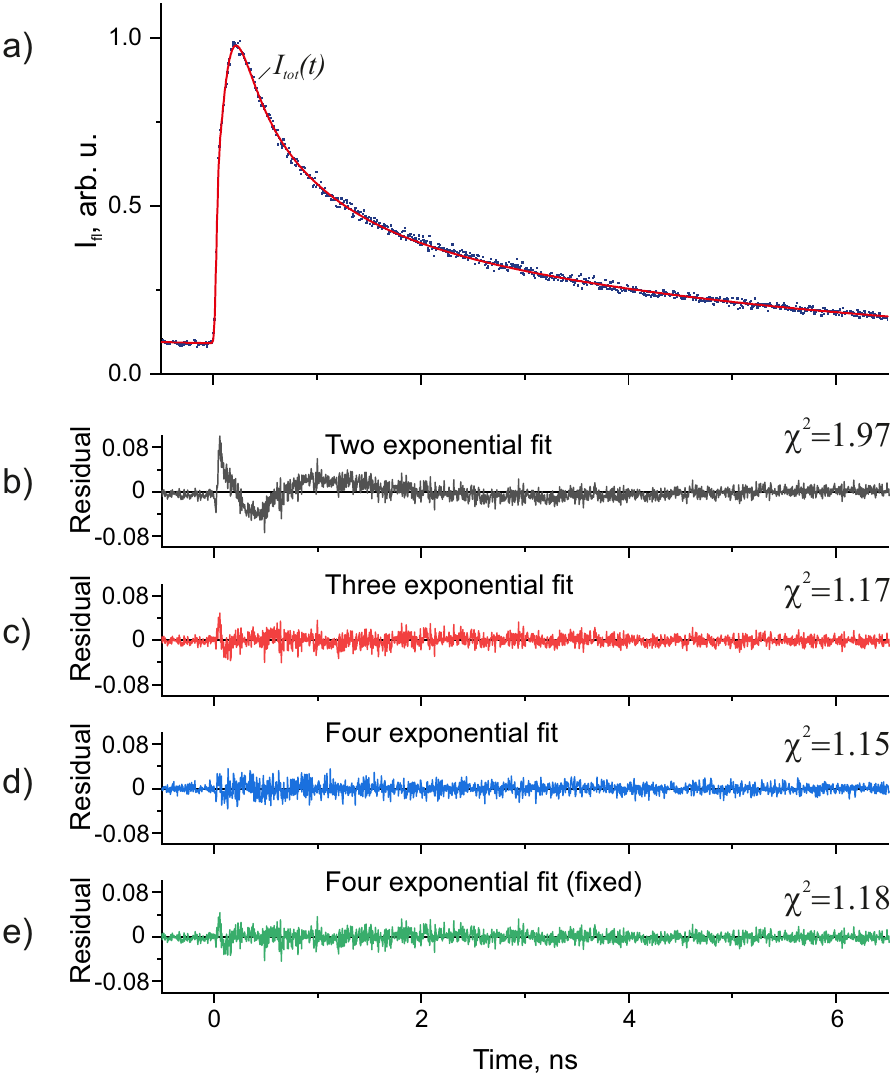}
   \end{tabular}
   \end{center}
   \caption[]
   { Isotropic fluorescence signal in NADH-ADH binary complex in solution and residuals related to various fitting models. \\
    a) The  fluorescence signal. Blue dots are experimental data and a red solid line is the best fit. \\
    b)--e) Residuals calculated using two-, three, and four-exponential fitting models.}
    \label{iso_binary}
\end{figure}

The polarization-insensitive fluorescence decay component $I_{tot}(t)$ obtained from two orthogonally polarized fluorescence components in Fig.~\ref{yx_curves} using eq.~(\ref{iso_func}) is shown in Fig.~\ref{iso_binary}a. The signal $I_{tot}(t)$ was analyzed using the multiexponential function in eq.~(\ref{sum_exp}). The best fit and residuals calculated using two-, three-, and four-exponential models are given in Fig.~\ref{iso_binary}b--\ref{iso_binary}e.  The obtained decay times $\tau_i$ and corresponding weighting coefficients $a_i$ are presented in Table~\ref{f_b_parameters}, where the fluorescence parameters given in the fourth line relate to the 4-exponential model were two decay times:  $\tau_{f2}$ = 0.24 and $\tau_{f3}$ = 0.66~ns were taken equal to those determined for free NADH in PBS buffer solution and fixed, while all other parameters were calculated from fit. The accuracy of determination  of all fluorescence parameters shown in Table~\ref{f_b_parameters} was less than 10\%.    

As can be seen all four models in Table~\ref{f_b_parameters} manifest significant contributions from the decay time $\tau_4$ with the value in the range of 3.65--4.85~ns that exceeds the decay times in free NADH in solution~\cite{Visser1981,Couprie1994a,Hull2001,Blacker2013,Blacker2019,Gorbunova20c} by about an order of magnitude and from the decay time $\tau_1$ of about 0.1~ns that is several times shorter than the decay times observed in free NADH.~\cite{Visser1981,Couprie1994a,Hull2001,Blacker2013,Blacker2019,Gorbunova20c}

The relatively long decay time of about 3--5~ns was observed earlier in ADH-containing solutions by several research groups~\cite{Gafni1976,Ladokhin1995,Konig1997,Piersma1998} and was associated with the ADH-NADH binary complex. The decay time  $\tau_4$ values in Table~\ref{f_b_parameters} are in perfect agreement with the results reported by Gafni et. al.~\cite{Gafni1976} (4.2~ns) and by Ladokhin and Brand~\cite{Ladokhin1995} (4.8~ns), however they are somehow larger than that reported by K\"onig et. al.~\cite{Konig1997} (2.9~ns) and by Piersma et. al.~\cite{Piersma1998} (3.16~ns).

The existence  of the short decay time  $\tau_1\approx$ 0.1~ns shown in the first column in Table~\ref{f_b_parameters} in enzyme-bound NADH was reported earlier by Ladokhin and Brand~\cite{Ladokhin1995} in ADH, by Piersma et al.~\cite{Piersma1998} in Nicotinoprotein Alcohol Dehydrogenase (np-ADH), and by  Vishwasrao et al.~\cite{Vishwasrao2005} in solutions and cells and was associated with fast relaxation process in enzyme-NADH boundary complex.

The values $N_f$ and $N_b$ in two last columns in  Table~\ref{f_b_parameters} are relative ground state concentrations of free and bound NADH calculated according to eqs.~(\ref{Nf}) and (\ref{Nb}) below. They are presented in the units of the constant $C$ given by:
   \begin{equation}
\label{C}
 C=\frac{\tau_{rad}}{\mathcal{E}_{pu}\mathnormal{l} \sigma},
\end{equation}
where $\tau_{rad}$ is a radiation decay time of the NADH excited state, $\mathcal{E}_{pu}$ is a laser pulse energy, $\mathnormal{l}$ is an absorption area length, and  $\sigma$ is an absorption cross section. 

The double exponential fit approximation presented in the first line in Table~\ref{f_b_parameters} is widely used for analysis of free and enzyme-bound NADH concentrations in solutions and cells~\cite{Gafni1976,Kierdaszuk1996,Konig1997,Vishwasrao2005,Skala2007,Vergen2012,Sharick2018,Schaefer2019}. Within this approximation the decay time $\tau_3$ in the first line in Table~\ref{f_b_parameters} can be associated with a mean fluorescence decay time in free NADH. As can be seen this approximation was not adequate in the conditions of our experiment in solution where the signal-to-noise ratio was relatively high because of the unappropriate residual in Fig.~\ref{iso_binary}b and large $\chi^2$ value of 1.97 in Table~\ref{f_b_parameters}.

The triple exponential approximation is presented in the second line in Table~\ref{f_b_parameters}. In this case the $\chi^2$ value and residual in Fig.~\ref{iso_binary}c are quite satisfactory. The decay time $\tau_3=0.77$~ns in the second line in Table~\ref{f_b_parameters} can be associated with the longest of two decay times observed in free NADH in buffer solution~\cite{Visser1981,Couprie1994a,Hull2001,Blacker2013,Blacker2019,Gorbunova20c}. We believe that the decay time $\tau_2=0.12$~ns in the second line in Table~\ref{f_b_parameters} is in fact an average of the short  decay time in free NADH of about 0.3~ns~\cite{Visser1981,Couprie1994a,Hull2001,Blacker2013,Blacker2019,Gorbunova20c} and the short decay time in the ADH-NADH binary complex of $\tau_1$=0.1~ns shown in the third and fourth lines in Table~\ref{f_b_parameters}.

\begin{table}[h]
 \centering \caption{Isotropic fluorescence decay parameters in ADH+NADH solution.}
 \label{f_b_parameters}
\begin{tabular}{cccccccc}
    \hline
      & $\tau_{1}$, ns ($a_1$) & $\tau_{2}$, ns ($a_2$) & $\tau_{3}$, ns ($a_3$) & $\tau_{4}$, ns ($a_4$) & $\chi^2$ & $N_f$ & $N_b$ \\
    \hline
  2 exp & - &  - & 0.30 (0.63) & 3.65 (0.37) & 1.97 & 0.63 & 0.37 \\
  3 exp &  - & 0.12 (0.50) & 0.77 (0.25) & 4.69 (0.25) & 1.17 & 0.25 & 0.26 \\
  4 exp & 0.10 (0.38) & 0.30 (0.24) & 0.90 (0.17) & 4.85 (0.21) & 1.15 & 0.19  &  0.21 \\
  4 exp (fix) & 0.10 (0.45) & 0.24 (0.05) & 0.66 (0.25) & 4.50 (0.25) & 1.18 & 0.20 & 0.25 \\
  \hline
\end{tabular}
\end{table}

The third line in Table~\ref{f_b_parameters} represents the results of the four-exponential fit approximation of the isotropic fluorescence signal in Fig.~\ref{iso_binary}a. Here the decay times $\tau_2$ and $\tau_3$ can be associated with the short and long lifetimes in free NADH and the decay times $\tau_1$ and $\tau_4$ can be associated with the short and long lifetimes in ADH-bound NADH. All four time values are in qualitative agreement with the decay times reported for free and bound NADH forms elsewhere~\cite{Lakowicz1996,Yu2009,Yaseen2013,Evers2018,Yaseen2017}.

For reducing of the number of fitting parameters and simplifying the fitting procedure the four-exponential fit was also  performed with the decay times  $\tau_{2}$ = 0.24~ns and $\tau_{3}$ = 0.66~ns fixed at their values  determined in free NADH in buffer solution~\cite{Gorbunova20c}, while all other parameters were calculated from fit. The obtained results are given in the fourth line in Table~\ref{f_b_parameters}.  As can be seen in this case the $\chi^2$ value and residual in Fig.~\ref{iso_binary}e are quite satisfactory and also the decay times agree qualitatively with those reported elsewhere~\cite{Lakowicz1996,Yu2009,Yaseen2013,Evers2018,Yaseen2017}. However, the weighting coefficients $a_2$ and $a_3$ and  their ratio are not in agreement with those determined in free NADH in aqueous solution.~\cite{Krishnamoorthy1987,Blacker2019,Sasin19,Vasyutinskii2017,Gorbunova20c}

In general, each of the decay times $\tau_i$ in Table~\ref{f_b_parameters} can be presented as:
 \begin{equation}
\label{gamma}
	\tau^{-1} = \tau^{-1}_{rad}+\tau^{-1}_{nrad},
\end{equation}
where $\tau_{rad}$ and $\tau_{nrad}$ are radiation and radiationless decay times, respectively.

The fluorescence quantum yield $Q$ is given by the expression:
  \begin{equation}
\label{Q}
	Q = \frac{\tau}{\tau_{rad}}.
\end{equation}

According to the experiments of Scott et al.~\cite{Scott1970} the fluorescence quantum yield of NADH in water was determined to be equal to $Q$=0.019 manifesting that the measured fluorescence decay rates in eq.~(\ref{gamma}) is dominated by the radiationless decay.

A common way for determination of the fractional population of free and bound species from the multiexponential expansion in eq.~(\ref{sum_exp}) is to assume that the radiation decay rate $\gamma_{rad}=\tau_{rad}^{-1}$ remains constant and then to suggest that the weighting coefficients $a_i$ in eq.~(\ref{sum_exp}) is equal to the fractional concentration of the species associated with it and to use the fraction of the total fluorescence signal generated by each species $i$~\cite{Vishwasrao2005,Sharick2018,Niesner2004}:
\begin{equation}
\label{f}
f=\frac{a_i\tau_i}{\sum_j a_j\tau_j }.
\end{equation}

However, as pointed out by Niesner et al.~\cite{Niesner2004} in general $a_i$ and $f$ values are influenced by the concentration, absorption cross-section, and fluorescence quantum yield. Consequently, it is necessary to know the photophysical properties of the fluorescing species in order to calculate their relative concentration.

For determination of the relative concentrations of bound and free NADH given in two last columns in Table~\ref{f_b_parameters} we used the following simplified model of absorption and emission of light by NADH. Integrating eq.~(\ref{sum_exp}) over time one obtains a well known expression for the averaged fluorescence intensity $I_{av}$ detected after each excitation laser pulse:~\cite{Lakowicz1996}
\begin{equation}
\label{FlTot}
I_{av}=\sum_i a_i\tau_i,
\end{equation}
where $a_i\tau_i$ is a contribution to the averaged fluorescence intensity from the species $i$ that can be presented in the form:
  \begin{equation}
\label{ai}
 a_i\tau_i= \mathcal{E}_{pu} \mathnormal{l} N_i\sigma_iQ_i,
\end{equation}
where $N_i$ is the ground state concentration and $Q_i$ is the quantum yield of the species $i$, respectively.

Assuming that  the radiative decay time $\tau_{rad}$ and the absorption cross section $\sigma$ remain constant all fluorescence parameters in Table~\ref{f_b_parameters} were referred to either bound, or free NADH forms. As shown above the decay times $\tau_2$ and $\tau_3$ in the third and fourth lines in Table~\ref{f_b_parameters} can be associated with the short and long lifetimes in free NADH and the decay times $\tau_1$ and $\tau_4$ can be associated with the short and long lifetimes in ADH-bound NADH. 

Substituting eq.~(\ref{Q}) and eq.(\ref{ai}) into eq.(\ref{FlTot}) the concentrations $N_f$ and $N_b$ can be readily written as:
 \begin{equation}
\label{Nf}
 N_f=C\frac{(a_2\tau_2+a_3\tau_3)}{(\tau_2+\tau_3)},
\end{equation}
   \begin{equation}
\label{Nb}
 N_b=C\frac{(a_1\tau_1+a_4\tau_4)}{(\tau_1+\tau_4)},
\end{equation}
where the constant $C$ is given in eq.~(\ref{C}).

As can be seen in Table~\ref{f_b_parameters} all three- and four- exponential models give similar values of $N_f$ and $N_b$. The two-exponential model gives the concentration $N_b$ that is in some agreement with other models in Table~\ref{f_b_parameters}, however the ratio $N_f/N_b$ given by this model differs from that given by the others.

Therefore, as can be seen in Table \ref{f_b_parameters} and Fig. \ref{iso_binary} the four-exponential fit of the isotropic fluorescence signals in lines 3 and 4 in  Table~\ref{f_b_parameters} provides the best characterization of free and enzyme-bound forms of NADH in solutions needed for detailed understanding of the relaxation processes occurring. The  four-exponential fit with two fixed free NADH lifetimes is simpler and results in practically the same fluorescence parameters for bound NADH, however gives incorrect ratio of the weighting coefficients $a_2$ and $a_3$ for free NADH. The three-exponential fit is also simpler and provides practically the same quality and plausibility of the fluorescence parameters for bound NADH. 

Note that for determination of only the ground state concentrations of enzyme-bound and free forms of NADH in cells that are of the main  interest for biochemical applications, the three-exponential fit looks preferable as it is simpler and can be used without practically any loss of accuracy. The two-exponential fit was shown to give an incorrect ratio $N_f/N_b$  in the conditions of our experiment in solution where the signal-to-noise ratio was relatively high.

\subsection{Excited state lifetime enhancement and heterogeneity in enzyme-bound NADH }
 \label{sec:heterogeneity}

In aqueous solution free NADH is known to exhibit biexponential fluorescence decay with the lifetimes of about 0.3~ns and 0.7~ns~\cite{Visser1981,Couprie1994a,Hull2001,Blacker2013,Blacker2019,Gorbunova20c}, while in enzyme-containing solutions and in cells it is widely used to fit NADH fluorescence decay by two exponentials, one of them having the decay times of a few nanoseconds and another of about 0.4~ns. These two decay times  are usually attributed to the enzyme-bound and free forms, respectively~\cite{Gafni1976,Kierdaszuk1996,Konig1997,Vishwasrao2005,Skala2007,Vergen2012,Sharick2018,Schaefer2019}, where the decay time of the enzyme-bound form  is about an order of magnitude larger than that in the free form. A significant enhancement of the decay time in enzyme-bound NADH was usually attributed with low polarity of the corresponding enzyme binding site~\cite{Piersma1998} however to the best of our knowledge a more detailed explanation was so far not available.

A  quantitative quantum chemical treatment of the fluorescence decay times in free and enzyme-bound  polyatomic molecules like NADH including the role of radiation and radiationless decay rate channels is  a challenging task that still needs to be carried out. In this paper we developed a simple model for qualitative explanation of the behavior of the decay rates observed in free and enzyme bound NADH.

In our recent paper~\cite{Gorbunova20c} the heterogeneity in the measured decay times in free NADH in water-methanol solutions was explained through \emph{ab initio} calculations that indicated different charge distributions in the \emph{cis} and \emph{trans} configurations of the NA ring. This finding suggested that \emph{cis} and \emph{trans} configurations are likely responsible for the heterogeneity in the measured decay times because different charge distributions  result in different intramolecular electrostatic field distributions and lead to different radiationless decay rates. The statement was based on the recent publication of Nakabayashi \emph{et al.} \cite{Nakabayashi2014} who studied the influence of external electric field on the fluorescence decay rate in NADH excited state and demonstrated that application of the electric field increased the probability of non-radiative decay. The increase of non-radiative decay rate was attributed to the $\pi\pi^*$ character of the electronic transition of the nicotinamide moiety.\cite{Nakabayashi2014} Moreover, Nakabayashi \emph{et al.} \cite{Nakabayashi2014} reported a significant shortening of the fluorescence mean decay time in NADH with increasing of the solution polarity.

Particular routes of the radiationless decay in NADH are still unknown. These can be either interactions of excited NADH molecules with environment, or intramolecular internal conversion to the ground electronic state, or intersystem crossing to the triplet manifold.~\cite{Lakowicz2006}  An  important (but probably not the only one) decay mechanism is the $\pi$-stacking interaction between NA and adenine moieties.~\cite{Liang2013}   The importance of this mechanism is supported by a significant increase of the measured NADH quantum yield upon increasing of the methanol concentration in water-methanol solution due to the growing of the contribution from unfolded NADH conformations.

\begin{figure} [h]
   \begin{center}
   \begin{tabular}{c}
   \includegraphics[height=7cm]{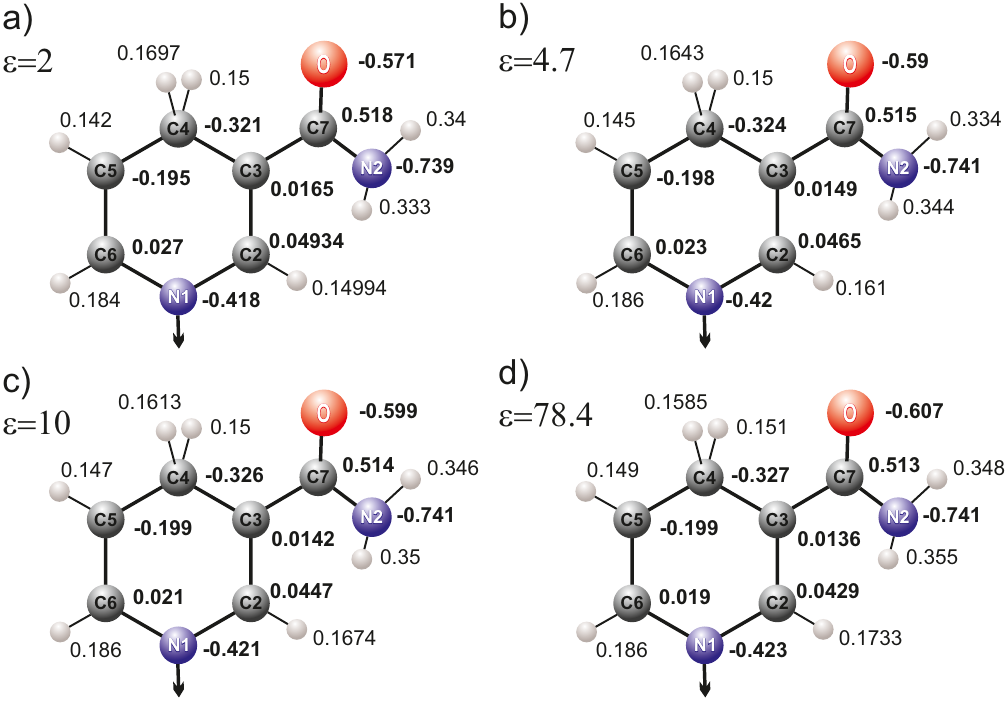}
   \end{tabular}
   \end{center}
   \caption{Charge distributions within NA ring calculated at various solution dielectric constant values $\varepsilon$  for the fixed NADH  unfolded \emph{trans} nuclear conformation shown in Fig.~\ref{fig:ADH-NADH}.  }
    \label{fig:NA}
\end{figure}

For extending the above suggestion to the case of ADH-bound NADH  we performed \emph{ab initio} calculations of electronic structure of the NADH conformer (PDB ID: 4XD2) shown in Fig.~\ref{fig:ADH-NADH}  in solutions with various polarities that modelled different enzyme binding sites. Calculated charge distributions (in Mulliken charges) within the NA ring related to the dielectric constant values of $\varepsilon$ = 2, 4.7, 10, and 78.4 are given in Fig.~\ref{fig:NA}.

\begin{figure} [ht]
   \begin{center}
   \begin{tabular}{c} 
   \includegraphics[height=9cm]{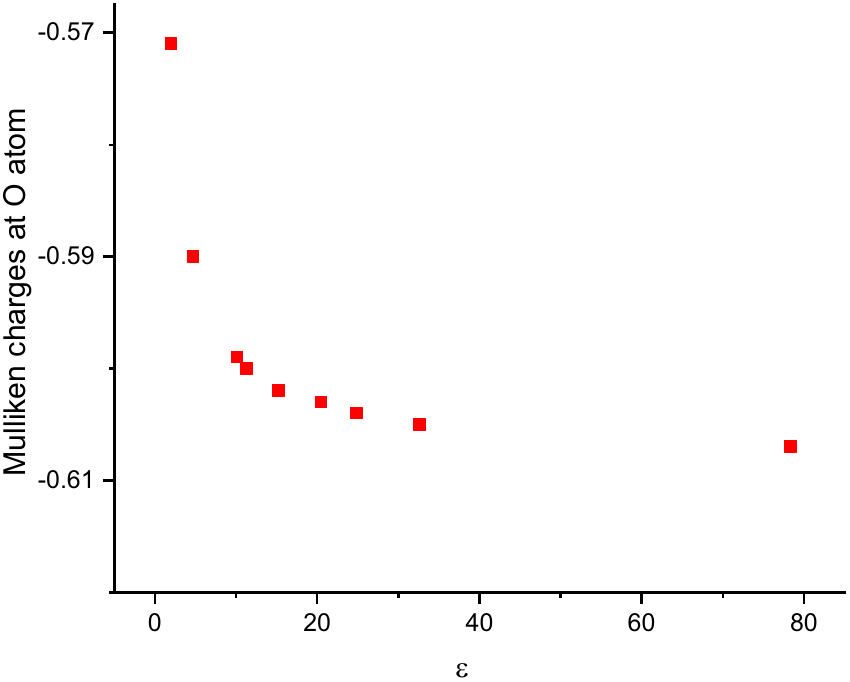}
   \end{tabular}
   \end{center}
   \caption{Charge on atom O in NA ring in Fig.~\ref{fig:NA}}
   \label{fig:O}
\end{figure}

Differences between charge distributions can be clearly seen in  Fig.~\ref{fig:NA} especially at relatively small $\varepsilon$ values.  For example, the charge on the oxygen atom is given in Fig.~\ref{fig:O} as a function of the dielectric constant values $\varepsilon$, where $\varepsilon$=78 relates to water and $\varepsilon$=2 is close to that in cyclohexane. As can be seen in Fig.~\ref{fig:O} at $\varepsilon \leq$10 the negative charge on the oxygen atom exhibits a sharp decrease dropping down by about 5\%. Having in mind that intramolecular electrostatic fields are typically very strong, of the order of 10$^5$V/cm, the charge changes at small $\varepsilon$ in Fig.~\ref{fig:O} can significantly affect the radiationless decay rates and result in dramatic increase of the measured lifetimes in enzyme-bound NADH. The model above suggests that the measured fluorescence lifetime depends mostly on the charge distributions in the NA ring and on the corresponding electrostatic field distribution. Therefore, the measured enhancement of the lifetime value in enzyme-bounded NADH can be attributed to a significant decrease of the charges separation in the NA ring in the conditions of a typical apolar binding site environment in intracellular NADH-linked dehydrogenase.~\cite{Piersma1998,Vishwasrao2005}

The above discussion is directly related to the NAD(P)H spectroscopy and redox imaging in cells that are now widely utilized since the pioneering works of Chance latin et al. ~\cite{Chance1962,Barlow1976} and Lackowicz et al.~\cite{Lakowicz1992}  The main two stereospecific conformers  that are widely discussed in the literature are \emph{syn} and \emph{anti} that can be transformed between each other simply by rotating the NA ring on 180$^{\circ}$ around the glycosidic bond~\cite{Vidal2018,Weinhold1991} that can be seen in Figs.~\ref{fig:ADH-NADH} and \ref{fig:NA} (in Fig.~\ref{fig:NA} the glycosidic bond is shown with  down arrows). 

It is important to note that both \emph{syn} and \emph{anti} conformers belong to the $trans$-type conformers of the NA ring~\cite{Plapp17}. As known from the X-ray structure spectroscopy and NMR experiments, \emph{anti} conformer is in favor by nature while the \emph{syn} conformer decreases the redox reactions rate~\cite{Hammen2002,Plapp17,Vidal2018}.  Therefore, the existence of a single nanosecond decay time attributed to the enzyme-bound NADH~\cite{Gafni1976,Kierdaszuk1996,Konig1997,Vishwasrao2005,Skala2007,Vergen2012,Sharick2018,Schaefer2019} can be understood within the above model because NAD(P)H imbedded into the ADH binding site is always fixed in the \emph{anti} \emph{trans}-like conformation.

\subsection{Excited state anisotropy decay.}
 \label{sec:heterogeneity}

For complete analysis of the polarization sensitive fluorescence decay  signals in NADH-ADH solution  the orthogonal fluorescence components $I_{\perp}(t)$ and $I_{\parallel}$ shown in Fig.~\ref{yx_curves} were presented as a sum of contributions from enzyme-bound and free NADH:
\begin{eqnarray}
\label{func_global}
I_{\parallel}(t)&=&GI_0\Big[I^{(f)}_l(t)(1+2r_l^{(f)}(t)) + I^{(b)}_{l}(t)(1+2r_l^{(b)}(t))\Big]\ast \mathrm{IRF}(t),
\\
I_{\perp}(t) &=& I_0\Big[I_{l}^{(f)}(t)(1-r_l^{(f)}(t)) + I_{l}^{(b)}(t)(1-r_l^{(b)}(t))\Big]\ast \mathrm{IRF}(t),
\label{func_global1}
\end{eqnarray}
where  $I_{l}^{(f)}(t)$ and $I_{l}^{(b)}(t)$  are isotropic contributions to the experimental signal for free and bound NADH, respectively, $r_l^{(f)}(t)$ and $r_l^{(b)}(t)$  are anisotropies of free and bound NADH, respectively, and the symbol $\ast$ denotes convolution.

In earlier  experiments on NADH fluorescence decay in solutions and cells~\cite{Gafni1976,Konig1997,Visser1981,Couprie1994a,Hull2001,Vishwasrao2005,Vergen2012,Blacker2013,Sharick2018,Schaefer2019,Blacker2019,Gorbunova20b} the isotropic contribution to the fluorescence signal in free NADH  $I_{l}^{(f)}(t)$ in eqs.~(\ref{func_global}) and (\ref{func_global1}) was presented either as a sum of two exponentials with decay times $\tau_{f1}$ and $\tau_{f2}$ and weighting coefficients $a_{f1}$ and $a_{f2}$:
 \begin{equation}
\label{If}
	I_{l}^{(f)}(t) = a_{f1}e^{-t/\tau_{f1}}+a_{f2}e^{-t/\tau_{f2}},
\end{equation}
or in a single exponential form:
 \begin{equation}
\label{If1}
	I_{l}^{(f)}(t) = a_{f}e^{-t/\tau_{f}},
\end{equation}
where the decay time $\tau_{f}$ was assumed to be an average of the decay times $\tau_{f1}$ and $\tau_{f2}$ in eq.~(\ref{If}).

The anisotropic contribution of free NADH $r_l^{(f)}(t)$ was usually taken in the form of a single exponential function with a rotational diffusion time $\tau_{fr}$ and an anisotropy $r_{f}$:
 \begin{equation}
\label{rf}
	r_l^{(f)}(t) = r_{f}e^{-t/\tau_{fr}}.
\end{equation}

According to the results of Sec.~\ref{sec:nonfluores} within the Model I that takes into account a reversible reaction between an enzyme-bound fluorescence probe and non-fluorescing amino acid the isotropic intensity $I_{l}^{(b)}(t)$ in bound NADH is given by a double exponential form in eq.~(\ref{eq:I(t)nsym}):
 \begin{equation}
\label{Ib}
	I_{l}^{(b)}(t) = a_{b1}e^{-t/\tau_{b1}}+a_{b2}e^{-t/\tau_{b2}}
\end{equation}
and the anisotropy $r_l^{(b)}(t)$ can be presented in the form of~eqs.~(\ref{eq:rij}), (\ref{eq:r11}):
\begin{equation}
\label{eq:rb11}
	r_l^{(b)}(t) = \Big(r_{b1} + r_{b2}e^{-t/\tau_{bv}}\Big)e^{-t/\tau_{br}},
\end{equation}

The rotational diffusion time of a protein as a whole was estimated to be about 30--40~ns~\cite{Piersma1998,Vishwasrao2005,Yu2009} that is much longer than other decay times observed in NADH. It is usually very inconvenient to measure such a long decay time using the TCSPC technique. Therefore, under the conditions of our experiment the term $e^{-t/\tau_{br}}$ in eq.~(\ref{eq:rb11}) was set to unity.

As shown in Sec.~\ref{sec:identical} within the Model II that takes into account exchange interaction between two identical enzyme-bound fluorescence probes the isotropic intensity $I_{iso}^{(b)}(t)$ shown in eq.~(\ref{eq:I(t)sym}) exhibits a single exponential form and the anisotropy $r^{(b)}(t)$ in eq.(\ref{eq:r(t)sym}) can be presented in the form of eq.~(\ref{eq:rb11}), where $1/\tau_{rb}=2W$.

The analysis of the polarization-insensitive component of experimental fluorescence signals in Sec.~\ref{sec:isotropic} suggests that the ADH-bound NADH is characterised by two decay times: $\tau_1$ of about 0.1~ns and $\tau_4$ of about 4.5~ns. Therefore the Model II in the form discussed in Sec.~\ref{sec:identical} cannot be directly applied for interpretation of the results of this paper  experiments. For the same reason the Model III is inappropriate for interpretation of the results.  

However, if two NADH molecules bound with the same enzyme dimer are contained not in the same micro environmental conditions (for instance, if the coefficients $\gamma_1$ and $\gamma_2$ in eqs.~(\ref{eq:master}) and (\ref{eq:master1}) are not equal to each other) the solution of the master equations eqs.~(\ref{eq:master}) and (\ref{eq:master1}) would contain a double exponential decay. Also, if interactions of NADH with another NADH and with a nonfluorescing amino acid occur simultaneously the Models I and II mix with each other and double exponential decay of the isotropic intensity $I_{iso}^{(b)}(t)$ becomes possible. We concluded that the analysis of our experimental data could not allow to distinguish between the Models I and II. Therefore in the following the isotropic intensity $I_{l}^{(b)}(t)$ and the anisotropy $r_{l}^{(b)}(t)$ of bound NADH are used in the double exponential forms in eqs.~(\ref{Ib}) and (\ref{eq:rb11}) for analysis of the anisotropic contributions $r^{(b)}(t)$ to the fluorescence decay in ADH-bound NADH shown in Fig.~\ref{yx_curves}.

The fluorescence parameters obtained by global fit of eqs.~(\ref{func_global}) and (\ref{func_global1}) using the isotropic fluorescence signal in free NADH in eq.~(\ref{If}), anisotropy in free NADH in eq.~(\ref{rf}) and the corresponding contributions from the bound NADH in eqs.~(\ref{Ib}) and (\ref{eq:rb11}) are presented in Table~\ref{tab:model1}. Within the fitting procedure the decay times $\tau_{f1}$, $\tau_{f2}$, and the anisotropy $r_{f}$ were fixed to their values determined in free NADH in aqueous solution~\cite{Couprie1994a,Vasyutinskii2017,Gorbunova20c}: $\tau_{f1}$ = 240~ps, $\tau_{f2}$ = 660 ps,  and $r_{f}$ = 0.49, respectively. The sum of the weighting coefficients in Table~\ref{tab:model1} was normalized to unity: $a_{f1}+a_{f2}+a_{b1}+a_{b2}=1$.

As can be seen in Table~\ref{tab:model1} all weighting coefficients $a_{f1}, a_{f2}, a_{b1},  a_{b2}$ and decay times $\tau_{b1}$, $\tau_{b2}$  are in perfect agreement with the corresponding values obtained through the analysis of polarization-insensitive fluorescence intensity in Sec.~\ref{sec:isotropic} and given in the last line in Table~\ref{f_b_parameters}.

\begin{table}[h]
\centering \caption{Fluorescence decay parameters of binary complex ADH-NADH calculated by global fit from eqs.~(\ref{func_global}), (\ref{func_global1}) using eqs.~(\ref{If}), (\ref{rf})--(\ref{eq:rb11}). \\    The fluorescence decay times $\tau_{f1}$ = 240 ps and $\tau_{f2}$ = 660 ps  and anisotropy $r_{f}$ = 0.49 in free NADH were fixed.}
\begin{tabular}{c c c c c c c c c c c c }
    \hline
       [NADH]:[ADH] & $\tau_{f1},$ ns &  $\tau_{f2},$ ns &  $\tau_{b1},$ ns & $\tau_{b2},$ ns&  $\tau_{fr},$ ns& $r_{b1}$ & $r_{b2}$ & $\tau_{bv},$ ns & $\chi^2$  \\
       $\mu M$ &  $(a_{f1})$ &  $(a_{f2})$ &  $(a_{b1})$ & $(a_{b2})$ &  &   & &  &   & \\
    \hline
 25 : 50  & 0.24 & 0.66  &  0.09  &  4.43  &  0.17 & 0.05 & 0.26 & 0.89  &  1.08   \\
  & (0.05) & (0.25) & (0.43) &  (0.27) &   &  &  &   &  \\ \hline
\end{tabular}
\label{tab:model1}
\end{table}

Also, Table~\ref{tab:model1} contains the terms describing anisotropic interactions of NADH with polarized light. The rotational diffusion time in free NADH, $\tau_{fr}$=0.17~ns in Table~\ref{tab:model1} agrees perfectly with that determined in aqueous solution (0.18~ns)~\cite{Couprie1994a,Vasyutinskii2017,Gorbunova20c}.  The vibrational relaxation time in bound NADH, $\tau_{bv}$ is discussed later.

\begin{table}[h]
 \centering \caption{Fluorescence decay parameters of binary complex ADH-NADH calculated by global fit from eqs.~(\ref{func_global}), (\ref{func_global1}) using eqs.~(\ref{If1})--(\ref{eq:rb11}). \\    The fluorescence decay times $\tau_{f1}$ = 240 ps and $\tau_{f2}$ = 660 ps  and anisotropy $r_{f}$ = 0.49 in free NADH were fixed.}
\begin{tabular}{ccccccccc}
    \hline
     [NADH]:[ADH] & $\tau_f$, ns  & $\tau_{b1}$, ns &  $\tau_{b2}$, ns & $\tau_{rf}$, ns & $r_{b1}$ & $r_{b2}$ & $\tau_{bv}$, ns & $\chi^2$  \\
     $\mu$M &  ($a_f$) &  ($a_{b1}$) & ($a_{b2}$) &  & &  &  &   \\
    \hline
    25 : 50 &  0.45 & 0.07 & 4.01  & 0.17  &  0.05  &   0.26 &  0.83 &  1.13  \\
    &  (0.32)  &  (0.39)  &  (0.28)  &  &    &    &   &    \\
    \hline
\end{tabular}
\label{tab:model2}
\end{table}

The fluorescence parameters obtained by global fit of eqs.~(\ref{func_global}) and (\ref{func_global1}) using the isotropic fluorescence signal in free NADH in eq.~(\ref{If1}), anisotropy in free NADH in eq.~(\ref{rf}), and 
the corresponding contributions from the bound NADH in eqs.~(\ref{Ib}) and (\ref{eq:rb11}) are presented in Table~\ref{tab:model2}. Within the fitting procedure the averaged decay time $\tau_{f}$ and the anisotropy $r_{f}$  in free NADH were fixed to their values $\tau_{f}$ = 450~ps and $r_{f}$ = 0.49, respectively. As before, the sum of the weighting coefficients in Table~\ref{tab:model2} was normalized to unity: $a_{f}+a_{b1}+a_{b2}=1$. The accuracy of determination  of all fluorescence parameters shown in Tables~\ref{tab:model1} and \ref{tab:model2} was less than 12\%.

As can be seen practically all isotropic and anisotropic calculated fluorescence parameters given in Table~\ref{tab:model2} are in good agreement with those given in Table~\ref{tab:model1}. This result suggests that the single exponential form of the isotropic decay in free NADH in eq.~(\ref{If1}) resulted in the similar accuracy of the fluorescence parameter values in bound NADH  as the double exponential form in eq.~(\ref{If}).

\begin{figure} [ht]
   \begin{center}
   \begin{tabular}{c} 
   \includegraphics[height=9cm]{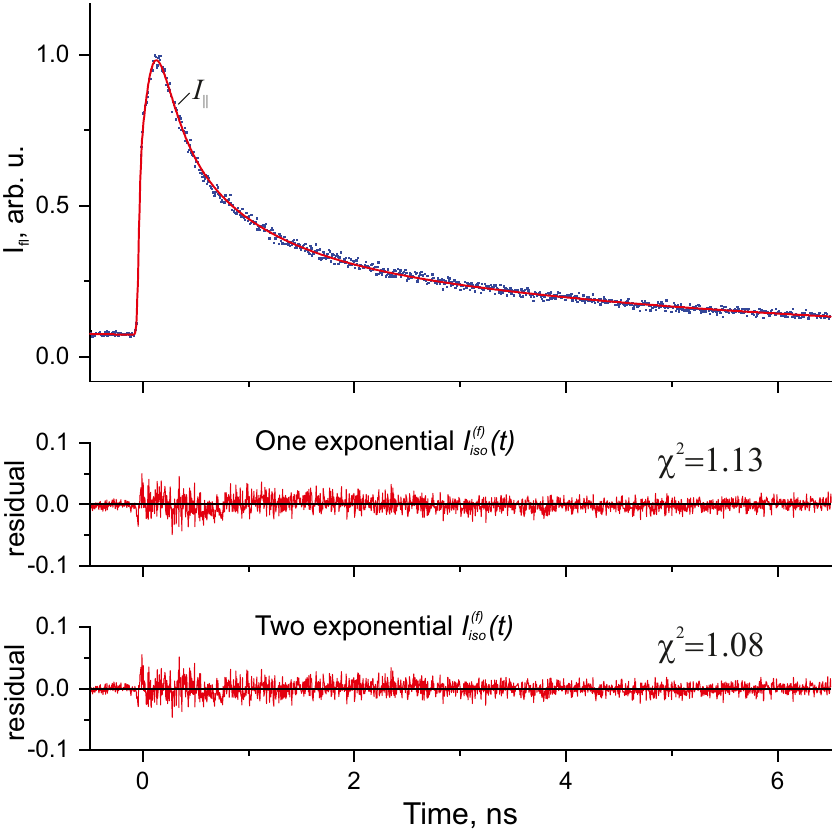}
   \end{tabular}
   \end{center}
   \caption{Fluorescence polarization component $I_{\parallel}(t)$ in ADH-NADH solution. \\
   Blue dots are experimental data, solid red curve is a fit, the residuals refer to the data in Tables~\ref{tab:model1} and \ref{tab:model2}, respectively. }
   \label{fig:parallel}
\end{figure}

The experimental fluorescence polarization component $I_{\parallel}(t)$ from Fig.~\ref{yx_curves} is presented in Fig.~\ref{fig:parallel} along with  fitting curves and residuals referred to three- and four exponential fit of the isotropic fluorescence signal $I_{iso}(t)=I_{iso}^{(f)}(t)+I_{iso}^{(b)}(t)$. As can be seen in Fig.~\ref{fig:parallel} the curves obtained within each of the fitting procedures and shown in Fig.~\ref{fig:parallel} practically coincide with each other. Although the corresponding residuals shown in Fig.~\ref{fig:parallel} and fitting parameter values given in Tabs.~\ref{tab:model1} and \ref{tab:model2} differ somehow from each other we can conclude that the use  of the averaged decay time $\tau_f$ for characterisation of free NADH instead of two decay times $\tau_{f1}$ and $\tau_{f2}$~\cite{Gafni1976,Kierdaszuk1996,Konig1997,Niesner2004,Vishwasrao2005,Skala2007,Vergen2012,Sharick2018,Schaefer2019} does not result in a significant loss of accuracy, although simplifies the fitting procedure.

A new feature of the data given in Tables~\ref{tab:model1} and \ref{tab:model2} is the anisotropic decay time $\tau_{bv}$=0.89~ns in the enzyme-bound NADH. To the best of our knowledge the sub-nanosecond anisotropic decay time in enzyme-bound NADH has never been observed before. We suggest that this decay time characterises fast anisotropic vibrational relaxation in excited states after the laser pulse that results in rotation of the NADH transition dipole moment due to rearrangement of nuclear configuration. The nuclear configurations and transition dipole moment directions in the NADH ground and relaxed excited states in water and methanol solutions were calculated \emph{ab initio} and reported in our recent publication.~\cite{Gorbunova20c}

Very likely that similar anisotropic vibrational relaxation time $\tau_v$ has recently been determined by Gorbunova et al.~\cite{Gorbunova20b} who studied  the dynamics of NADH excited states in water-ethanol solutions by means of a polarization pump-and-probe technique. As reported by Gorbunova et al.~\cite{Gorbunova20b} the decay time $\tau_v$ was about 2~ps in aqueous solution and then increased up to 20~ps in 80\% ethanol.  The significant increase of the decay time $\tau_{bv}$ in ADH-NADH complex observed in this paper can be attributed to the slowdown of vibrational relaxation under  the increase of effective viscosity and decrease of polarity in the binding site environment.

The existence of the anisotropic vibrational relaxation time $\tau_{bv}$ among the fluorescence parameters in Tables~\ref{tab:model1} and \ref{tab:model2} is important because it depends on the binding site viscosity and polarity and can be used together with other decay times for the binding site characterization. Note that this paper results do not support earlier suggestion of Vishwasrao et al.~\cite{Vishwasrao2005} who stated that the anisotropy-based approach offers the most sensitive discrimination of free and enzyme bound NADH. According to our analysis sub-nanosecond anisotropic decay times can be associated either with free, or with enzyme-bound NADH forms. In the former case this can be a contribution from free NADH rotational diffusion time while in the latter case a contribution from the  anisotropic vibrational relaxation time $\tau_{bv}$ and/or exchange time $1/2W$.

\section{Conslusions}
Theoretical and experimental studies of polarized fluorescence in NADH-ADH binary complexes in buffer solution have been carried out. A global fit procedure was used for determination of the fluorescence parameters from experimental data. The interpretation of the parameter values was supported by \emph{ab initio} calculations of NADH structure in solutions with various polarities. A quasiclassical model was developed for description of the polarized fluorescence decay in enzyme-NADH binary complexes taking into consideration reversible exchange reactions between two NADH molecules bounded within the same ADH dimer and energy transfer interactions between each NADH and surrounding amino acids. The results obtained suggest that the origin of a significant enhancement of the decay time in ADH-bounded NADH compare with free NADH can be due to the decrease of non-radiative relaxation probabilities resulting from the decrease of charges separations in the nicotinamide ring in the conditions of an apolar  ADH binding site environment.   Comparison between the experimental data of the fluorescence anisotropy decay in ADH-bound NADH under two-photon excitation with the quasiclassical model suggests the existence of an isotropic decay time of about 150 ps and an anisotropic decay time of about 1 ns both manifesting different interactions between NADH and ADH.  The analysis of the polarization-insensitive component of the fluorescence decay in ADH-containing solution has been carried out using various multiexponential fitting procedures.  The results obtained were compared with the results reported by other authors.

\section*{Acknowledgement}
IAG, MES, and OSV are grateful to Russian Foundation for Basic Research for financial support under the grant No 18-53-34001. IAG, AAS, YMB, and OSV are grateful to BASIS Foundation for financial support under
the grant no. 19-1-1-13-1. The authors are grateful to the Ioffe Institute for providing the equipment used in experiments.

\bibliography{Enzyme-NADH}

\end{document}